\documentclass[hyper,letterpaper,11pt]{article}
\pdfoutput=1\usepackage{a4wide}
\usepackage{CJK}
\usepackage{amsmath}
\usepackage[
      colorlinks=true,
      linkcolor=blue,
      urlcolor=blue,
      filecolor=black,
      citecolor=blue,
            ]{hyperref}
\usepackage{color}
\usepackage{graphicx}
\usepackage{amsfonts}
\usepackage{amssymb}
\usepackage{epstopdf}

\def\beq{\begin{equation}}
\def\eeq{\end{equation}}

\def\bc{\begin{center}}

\def\ec{\end{center}}
\def\be{\begin{eqnarray}}
\def\ee{\end{eqnarray}}

\date{}

\begin{document}

\begin{CJK*}{GBK}{song}

\title{\bf \Large Non-relativistic Josephson Junction from Holography}

\author{\large
~Huai-Fan Li $^1$\footnote{E-mail: huaifan.li@stu.xjtu.edu.cn}~,
~Li Li $^2$\footnote{E-mail: lili@physics.uoc.gr}~,
~Yong-Qiang Wang $^3$\footnote{E-mail: yqwang@lzu.edu.cn}~,
~Hai-Qing Zhang $^4$\footnote{E-mail: hqzhang@cfif.ist.utl.pt}\\
\\
\small $^1$ Institute of Theoretical Physics, Department of Physics,\\
\small Shanxi Datong University, Datong, 037009, China;\\
\small $^2$ Crete Center for Theoretical Physics, Department of Physics,\\
\small University of Crete, 71003 Heraklion, Greece;\\
\small $^3$ Institute of Theoretical Physics, Lanzhou University, Lanzhou, 730000, China;\\
\small $^4$ CFIF, Instituto Superior T\'ecnico, Universidade de Lisboa, \\
\small Av. Rovisco Pais, 1049-001 Lisboa, Portugal.}

\maketitle

\vspace{-10cm}\hspace{12cm}{CCTP-2014-23}
\vspace{10cm}

\vspace{-10cm}\hspace{11.7cm}{CCQCN-2014-48}
\vspace{10cm}

\begin{abstract}
\normalsize We construct a Josephson junction in non-relativistic case with a Lifshitz geometry as the dual gravity.  We investigate the effect of the Lifshitz scaling in comparison with its relativistic counterpart. The standard sinusoidal relation between the current and the phase difference is found for various Lifshitz scalings characterised by the dynamical critical exponent. We also find the exponential decreasing relation between the condensate of the scalar operator within the barrier at zero current and the width of the weak link, as well as the relation between the critical current and the width. Nevertheless, the coherence lengths obtained from two exponential decreasing relations generically have discrepancies for non-relativistic dual.
\end{abstract}

\tableofcontents

\section{Introduction}

Traditional condensed matter paradigms with weakly interacting quasiparticles are challenged by strongly correlated electron systems. One of the profound examples is the high temperature superconductivity. The basic idea of the BCS theory, like the weak-coupling mean field approximation and phonon-mediated electron pairing mechanism is no longer applied without modifications. Therefore, to develop new theoretical framework and concepts is desirable to understand those strongly coupled many-body systems. On the other hand, holography~\cite{Maldacena:1997re,Gubser:1998bc,Witten:1998qj}, as a framework to access the strongly coupled regime of quantum field theory by its gravity dual living in a spacetime with higher dimensionality, has been useful in addressing the physical properties of strongly interacted condensed matter systems, such as high $T_c$ cuprates and heavy feimions.  Within this context, models for unconventional superconductors have been widely studied holographically. The first holographic superconductor, known as Abelian-Higgs model, has been introduced in ref.~\cite{Hartnoll:2008vx} in terms of a charged scalar field in the bulk whose condensate corresponds to a s-wave superconducting order.  This gravity setup was soon generalised to holographic p-wave models~\cite{Gubser:2008wv,Aprile:2010ge,Cai:2013aca} and d-wave models~\cite{Chen:2010mk,Benini:2010pr}, see refs.~\cite{ Herzog:2009xv,Horowitz:2010gk, Musso:2014efa} for good reviews.~\footnote{Holographic superconductors have been studied usually in the absence of dynamical electromagnetic fields, thus in the limit in which they coincide with holographic superfluids.  The dynamics of the electromagnetic field is very relevant for, such as, the Meissner effect and the exponential damping of the magnetic field in vortices. The authors of ref.~\cite{Domenech:2010nf} explained for the first time how to introduce a dynamical gauge field in holographic superconductors. }

Josephson junctions possess very important features in both theoretical and practical fields of superconductivity. A typical Josephson junction consists of two superconductors separated by a week contact. Depending on the specimen of the constituent superconductors and the nature of the contact, there are various kinds of junctions. The contact can be a normal conductor, an insulator, or a narrow superconductor. The corresponding junctions are referred to as SNS, SIS and SS'S junctions, respectively. Moreover, the coupled superconductors can be of different types. The authors of ref.~\cite{Horowitz:2011dz} constructed a holographic SNS junction by using the simplest holographic superconductor~\cite{Hartnoll:2008vx}. This junction exhibits the standard relation between the current $J$ across the junction and the phase difference $\gamma$ of the condensate, i.e. $J=J_{max}\sin(\gamma)$. The dependence of the maximum current (or critical current) $J_{max}$ on the temperature and size of the junction also reproduces familiar results. Soon after, this setup has been generalised to other types of Josephson junctions~\cite{Wang:2011rva,Siani:2011uj,Wang:2011ri,Wang:2012yj,Rozali:2013pla} as well as superconducting quantum interference device (SQUID)~\cite{Cai:2013sua,Takeuchi:2013kra}. A distinctly different way to construct a holographic model of Josephson junctions based on designer multi-gravity has been proposed in ref.~\cite{Kiritsis:2011zq} in which Josephson junction arrays were discussed.~\footnote{Holographic Josephson junctions from D-branes have been considered in ref.~\cite{Domokos:2012rj} aiming at providing a geometrical picture for the holographic dual. Through this way non-Abelian Josephson junctions and AC Josephson effect have been naturally realized.}

The above studies focused on gravity duals with asymptotic AdS boundary, which indicates that the dual theory is a relativistic conformal field theory. However, there exist many scale-invariant systems without the Lorentz invariance especially near the critical points~\cite{Hartnoll:2009sz,Sachdev:2010ch}. In particular, the electrons in real materials are in general non-relativistic, thus it is natural to ask whether one can develop a similar model with non-relativistic kinematics. The Lifshitz geometry as a dual gravity is a very natural candidate to describe those non-relativistic theories. Lifshitz geometry is characterised by the so-called {\it dynamical critical exponent z} which  governs the anisotropy between spatial and temporal scaling $t\rightarrow\lambda^z t, \vec{x}\rightarrow\lambda\vec{x}$. The case $z=1$ is nothing but the usual relativistic scaling. The Lifshitz holography has been used to address various aspects of non-relativistic systems, such as strange metal transport~\cite{Hartnoll:2009ns,Lee:2010uy,Kim:2010zq}, thermalization~\cite{Keranen:2011xs}, (non-)Fermi liquid~\cite{Gursoy:2011gz,Fang:2012pw,Wu:2013xta}  and so on.

The purpose of the present work is to investigate the Josephson junction of the non-relativistic theory with the Lifshitz geometry as a dual gravity.  We aim at the effects due to the Lifshitz scaling in comparison with the relativistic case $z=1$. More specifically, we construct holographic junctions in the Lifshitz black branes with $z=1, 2$ and $3$. Following ref.~\cite{Horowitz:2011dz}, we  consider the Abelian-Higgs model for holographic superconductors with inhomogeneous boundary conditions breaking  translational invariance. This model typically require us to solve complicated coupled partial differential equations (PDEs). Taking advantage of the Chebyshev spectral methods to solve those PDEs numerically, we find that the famous sinusoidal relation between the current and the phase difference across the weak link do exist no matter what $z$ is. The condensate of the operator at zero current in the middle of the link has an exponential decreasing relation with respect to the width of the link $\ell$; Meanwhile, the critical current $J_{max}$ also has an exponential decreasing relation to $\ell$. From the above exponential decreasing relations, one can extract the coherence length $\xi$ independently. In relativistic cases~\cite{Horowitz:2011dz,Wang:2011rva,Siani:2011uj,Wang:2011ri,Wang:2012yj}, the value of the coherence length $\xi$ fitted from critical current and condensate is consistent to each other within acceptable errors. However, for general $z\neq1$, this result is violated. A typical example exhibiting this violation is the case with $z=3$.

The paper is organised in the following: In Section~\eqref{sect:back} we derive the equations of motions in the Lifshitz black brane background; We show our numerical technique for dealing with non-trivial boundary conditions and numerical results in Section~\eqref{sect:result}; In Section~\eqref{sect:conclusion}, we draw our conclusion and give some comments to the Lifshitz Josephson junction.


\section{The Gravity Setup}
\label{sect:back}
%
%
%

We adopt the black brane background in d+2 dimensional spacetime as~\cite{Taylor:2008tg}
\begin{equation}\label{metric}
ds^2=L^2\left(-r^{2z}f(r)dt^2+\frac{dr^2}{r^2 f(r)}+r^2\sum^{d}_{i=1} dx_i^2\right),\quad f(r)=1-\frac{r_0^{z+d}}{r^{z+d}},
\end{equation}
where $z$ is the {\it dynamical critical exponent}, $r_0$ is the radius of horizon, and $d$ is the spacial dimension of the boundary. The asymptotical Lifshitz boundary is located at $r\rightarrow\infty$. This geometry for $z=1$ is nothing but the AdS-Schwarzschild black brane, while it is a gravity dual with the Lifshitz scaling as $z>1$. The Hawking temperature of this black brane is
\begin{equation}\label{solution}
T=\frac{z+d}{4\pi}r_0^z.
\end{equation}

In the probe limit of the above background, we consider the model of a $U(1)$ gauge field $A_\mu$ coupled a charged scalar field $\psi$ . The corresponding action reads
\begin{equation}\label{action2}
S=\int d^{d+2} x\sqrt{-g}\left[-|\nabla\psi-iA\psi|^2-m^2|\psi|^2-\frac{1}{4}F_{\mu\nu}F^{\mu\nu}\right],
\end{equation}
 in which $F_{\mu\nu}$ is a $U(1)$ gauge field strength with $F_{\mu\nu}=\partial_\mu A_\nu-\partial_\nu A_\mu$. The equations of motions (EoMs) can be obtained from the above action and the background as
\be \label{eom1}
0&=&(\nabla_\mu-iA_\mu)(\nabla^\mu-iA^\mu)\psi-m^2\psi, \\
\label{eom2}\nabla_\nu F^{\nu\mu}&=&i[\psi^*(\nabla^\mu-iA^\mu)\psi-\psi(\nabla^\mu+iA^\mu)\psi^*].
\ee
We choose the ansatz of the fields as
\begin{equation}\label{ansatz}
\psi=|\psi|e^{i\varphi},\quad A=A_t dt+A_r dr+A_{x_1}dx_1,
\end{equation}
where $|\psi|,\varphi,A_t,A_r,A_{x_1}$ are all real functions of $r$ and $x_1$. We would like to work with the gauge-invariant combination $M_{\mu}=A_{\mu}-\partial_{\mu}\varphi$.

Substituting the Lifshitz black brane background~\eqref{metric} and the ansatz~\eqref{ansatz} into the EoMs ~\eqref{eom1} and \eqref{eom2}, we can obtain the following coupled PDEs:~\footnote{ For convenience, we will define $x\equiv x_1$ in the following context.}
\begin{subequations}
\be
\partial_r^2|\psi|+\frac{1}{r^4 f}\partial_x^2|\psi|+(\frac{d+z+1}{r}+\frac{f'}{f})\partial_r|\psi|+\frac{1}{r^2 f}(\frac{M_t^2}{r^{2z}f}-r^2 f M_r^2-\frac{M_x^2}{r^2}-L^2 m^2)|\psi|=0,  \label{pde1}\\
\partial_r M_r+\frac{1}{r^4 f}\partial_x M_x+\frac{2}{|\psi|}(M_r\partial_r|\psi|+\frac{M_x}{r^4 f}\partial_x|\psi|)+(\frac{d+z+1}{r}+\frac{f'}{f})M_r=0, \label{pde2}\\
\partial_r^2 M_t+\frac{1}{r^4 f}\partial_x^2 M_t+\frac{d-z+1}{r}\partial_r M_t-\frac{2L^2|\psi|^2}{r^2 f}M_t=0, \label{pde3}\\
\partial_x^2 M_r-\partial_x\partial_r M_x-2L^2 r^2 |\psi|^2 M_r=0, \label{pde4}\\
\partial_r^2 M_x-\partial_x\partial_r M_r+(\frac{f'}{f}+\frac{d+z-1}{r})(\partial_r M_x-\partial_x M_r)-\frac{2|\psi|^2}{L^2 r^2 f}M_x=0,\label{pde5}
\ee\label{pdes}
\end{subequations}
where a prime $'$ denotes the derivative with respect to $r$. It is clear that the phase function $\varphi$ has been absorbed into the gauge invariant quantity $M_\mu$. The second equation \eqref{pde2} is a constraint equation which can be obtained from the algebraic combinations of \eqref{pde4} and \eqref{pde5} as $2r^2|\psi|^2\times\text{Eq.}\eqref{pde2}+\partial_r[\text{Eq.}\eqref{pde4}]+\partial_r[\text{Eq.}\eqref{pde5}]+[f'/f+(d+z-1)/r]\times\text{Eq.}\eqref{pde4}\equiv0$.
Therefore, in fact there are four independent EoMs with four fields, {\it i.e.}, $|\psi|, M_t, M_r$ and $M_x$.

In order to solve the above coupled PDEs, we need to impose suitable boundary conditions. First, we demand the regularity of the fields at the horizon. Since the metric component $g_{tt}$ is zero at the horizon, the field $M_t$ should be vanishing at the horizon, while other fields are finite at the horizon.

Near the infinite boundary $r\to\infty$, the fields $|\psi|, M_r$ and $M_x$ have the following asymptotic expansions,
\begin{equation} \label{uv}
\begin{split}
|\psi|=&\frac{\psi^{(1)}(x)}{r^{(z+d-\sqrt{(z+d)^2+4m^2})/2}}+\frac{\psi^{(2)}(x)}{r^{(z+d+\sqrt{(z+d)^2+4m^2})/2}}+\mathcal{O}(\frac{1}{r^{(z+d+\sqrt{(z+d)^2+4m^2})/2+1}}), \\
M_r=&\frac{M_r^{(1)}(x)}{r^{d+z-1}}+\mathcal{O}(\frac{1}{r^{d+z}}), \\
M_x=&\nu(x)+\frac{J(x)}{r^{d+z-2}}+\mathcal{O}(\frac{1}{r^{d+z-1}}).
\end{split}
\end{equation}
 However, the asymptotic behaviour of $M_t$ is more sophisticated depending on the values of $z$ and $d$,
\begin{equation} \label{uvmt}
\begin{split}
M_t=&\rho(x)-\mu(x)\mathrm{log}(r)+\mathcal{O}(\frac{1}{r}), \quad \text{for} \quad (d-z=0), \\
M_t=&\mu(x)-\frac{\rho(x)}{r^{d-z}}+\mathcal{O}(\frac{1}{r^{d-z+1}}), \quad \text{for} \quad (d-z<0 \text{~or~} 0<d-z<2),\\
M_t=&\mu(x)-\frac{\rho(x)}{r^{2}}+\frac{\partial_x^2\mu(x)}{2 r^2}\mathrm{log}(r)+\mathcal{O}(\frac{1}{r^{3}}), \quad \text{for} \quad (d-z=2),\\
M_t=&\mu(x)-\frac{\rho(x)}{r^{d-z}}{+\frac{\partial^2_x\mu(x)}{2(d-z-2)r^2}}+\mathcal{O}(\frac{1}{r^{d-z+1}}),\quad \text{for}  \quad (d-z>2) .
\end{split}
\end{equation}
The conformal dimension of the scalar field $|\psi|$ is $\Delta_\pm=(z+d\pm\sqrt{(z+d)^2+4m^2})/2$. In the following, we focus on the case $\psi^{(1)}\equiv0$, which means there is no source term of the dual scalar operator. We will always regard $\mu$ as the chemical potential, although for $z>d$ it is not the largest mode near the boundary~\cite{Hartnoll:2009ns}. According to the holographic dictionary, the coefficients $\psi^{(2)}$, $\rho$, $\nu$ and $J$ correspond to the condensate of the dual scalar operator $\langle \mathcal{O}\rangle$, charge density, superfluid velocity and current in the boundary field theory, respectively.~\footnote{We also notice that there is a relation $\partial_x^2 M_r^{(1)}(x)+(d+z-2)\partial_xJ(x)=0$, which can be used to set $J$=const by imposing $\partial_x M_r^{(1)}=0$, in the numerical calculations in the next section. }  Furthermore, the gauge invariant phase difference $\gamma=\Delta \varphi-\int A_x$ across the weak contact can be recast as~\cite{Horowitz:2011dz}
\begin{equation}\label{gamma}
\gamma=-\int^{+\infty}_{-\infty}dx[\nu(x)-\nu(\pm\infty)].
\end{equation}
In order to mimic a SNS Josephson junction, we choose the profile of the chemical potential $\mu(x)$ similar to that in ref.~\cite{Horowitz:2011dz}, which is given by
\begin{equation}\label{mu}
\mu(x)=\mu_\infty\left\{1-\frac{1-\epsilon}{2\tanh(\frac{\ell}{2\sigma})}\left[\tanh(\frac{x+\frac{\ell}{2}}{\sigma})-\tanh(\frac{x-\frac{\ell}{2}}{\sigma})\right]\right\},
\end{equation}
where $\mu_\infty=\mu(+\infty)=\mu(-\infty)$ is the chemical potential at $x=\pm\infty$, while $\ell$, $\sigma$ and $\epsilon$ control the width, steepness and depth of the junction, respectively.

Note that the coupled PDEs~\eqref{pdes} exhibit the following scaling symmetry:
\begin{equation}\label{scal}
 t\rightarrow\lambda^z t,\quad x_i\rightarrow\lambda x_i,\quad r\rightarrow\frac{1}{\lambda}r,\quad M_t\rightarrow\frac{1}{\lambda^z}M_t,\quad M_x\rightarrow \frac{1}{\lambda} M_x,\quad M_r\rightarrow \lambda M_r,
\end{equation}
with $\lambda$ an arbitrary constant.
Following ref.~\cite{Horowitz:2011dz}, we define the critical temperature of the junction $T_c$  identical to the critical temperature of a homogenous superconductor with vanishing current.~\footnote{Lifshitz holographic superconductors in homogenous case have studied, for example, in refs.~\cite{Brynjolfsson:2009ct,Sin:2009wi,Bu:2012zzb,Lu:2013tza}.} Therefore, $T_c$ is proportional to $\mu_\infty=\mu(+\infty)=\mu(-\infty)$ corresponding to the scaling symmetry~\eqref{scal}:
\begin{equation}\label{tem1}
T_c=\frac{(z+d)r_0^z}{4\pi\mu_c}\mu(\infty),
\end{equation}
where $\mu_c$ is the critical chemical potential for a homogenous superconductor without current at temperature $T=\frac{z+d}{4\pi}r_0^z$. Inside the junction, $x\sim(-\frac{\ell}{2},\frac{\ell}{2})$, the effective critical temperature reads
\begin{equation}\label{tem2}
T_0=\frac{(z+d)r_0^z}{4\pi\mu_c}\mu(0).
\end{equation}
Therefore, for $T_0<T<T_c$, the in-between junction is in the normal metallic phase, while the region outside the junction is in the superconducting phase. It is in this way one models the SNS Josephson junction by holography.

\section{Numerical Results}
\label{sect:result}
We take advantage of the Chebyshev spectral methods~\cite{trefensen} to numerically solve the EoMs~\eqref{pde1}-\eqref{pde5}. We first set $r_0=1$ by using of the scaling symmetry~\eqref{scal}. For the convenience, we also make the coordinate transformation in the following way $u=1/r$ and $y=\tanh(\frac{x}{4\sigma})$, as well as
\be
|\psi|&\rightarrow &\frac{|\psi|}{r^{(z+d-\sqrt{(z+d)^2+4m^2})/2}},\\
M_r&\rightarrow&\frac{M_r}{r^{d+z-1}}.
\ee
In the following, we will consider the case with $d=2$, but it can be straightforwardly generalised to other dimensions.
Specifically, we choose the {\it dynamical critical exponent} $z$ as $z=1, 2$ and $3$. It is well-known that $z=1$ is no other than the relativistic dual while $z=2$ and $z=3$ are for the non-relativistic theories. Physically, we would like to investigate the properties of the Josephson junctions with the same conformal dimension of each dual scalar operator, hence we set $\Delta_+=3$ as we vary $z$. Therefore, in this sense the mass square are $m^2=0, -3$ and $-6$ with respect to $z=1, 2$ and $3$.

The values of critical chemical potential $\mu_c$ (or in the sense of the critical temperature $T_c$ explained above) for the homogeneous superconductors are $\mu_c\approx7.5877, 9.0445$ and $9.7667$ with respect to $(z,m^2)=(1,0), (2,-3)$ and $(3,-6)$. Therefore, we choose a unified chemical potential $\mu(x)$ for the junction with the parameters $\mu_\infty=10.5, \sigma=0.7$ and $\epsilon=0.7$. The profile of the chemical potential would satisfy the requirement of the Josephson junctions for the three cases.

Near the spacial boundary $x=\pm\infty$, we demand that all the fields are independent of $x$ because of the flat $\mu(x)$ near $|x|\to\infty$. There is also a symmetry of the fields when we flip the sign of $x\to -x$,
\be
|\psi|\to|\psi|,\quad M_t\to M_t,\quad M_r\to-M_r,\quad M_x\to M_x.
\ee
Therefore, $M_r$ is an odd function of $x$ while others are even. Thus we can set $M_r(x=0)=0$, and other fields have vanishing first order  derivative with respect to $x$ at $x=0$. From the scaling symmetry~\eqref{scal} and the UV asymptotic expansions~\eqref{uv} and~\eqref{uvmt}, it is easy to see that the quantities $J/T_c^{(1+z)/z}$ and $\langle \mathcal{O}\rangle/T_c^{3/z}$ are dimensionless.

\subsection{The case of $z=1$}
\begin{figure}[h]
\centering
   \includegraphics[scale=0.35]{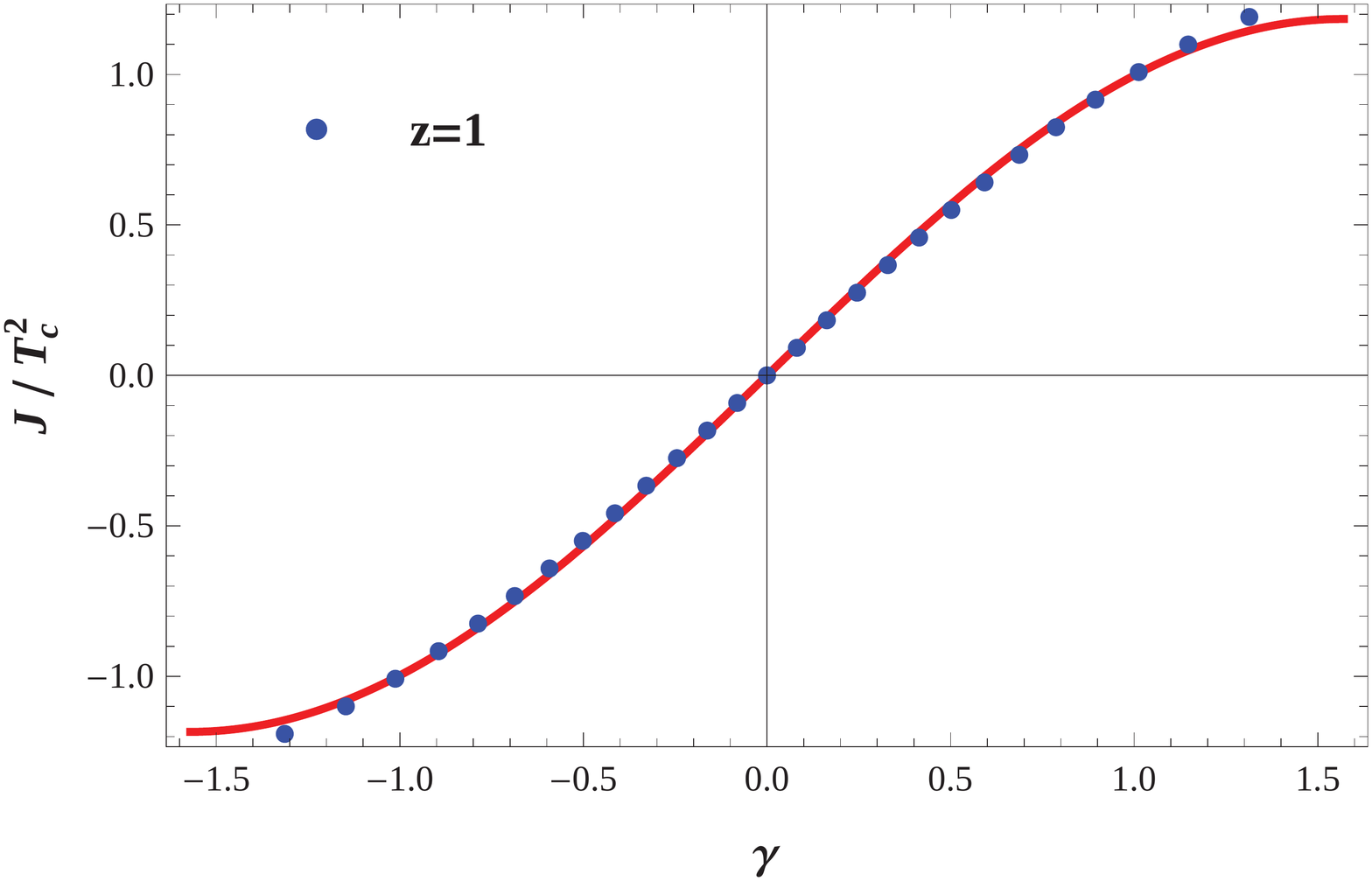}
      \caption{\label{Jgamma1} Relation between $J/T_c^{(1+z)/z}$ and $\gamma$ for $z=1$. The dots are from the numerics while the the red line is the fitted sin curves of these dots. We use $\mu_\infty=10.5, \ell=3, \sigma=0.7$ and $\epsilon=0.7$.  }
\end{figure}

For $z=1$, the asymptotic expansion of $M_t$ near the boundary is $M_t\sim\mu(x)-\frac{\rho(x)}{r}$.  It can be easily calculated as before~\cite{Horowitz:2011dz}. The relation between the current and the phase difference is shown in figure.~\eqref{Jgamma1}. The blue dots are for the data from numerical calculations while the red curve is fitted by the sinusoidal relation. We can read from the plots that
\be
J/T_c^2\approx 1.18436\sin(\gamma), \quad \text{for} \quad z=1.
\ee
For our choosing parameters in figure.~\eqref{Jgamma1} the critical current is $J_{max}/T_c^2\approx1.18436$.

It has been uncovered in refs.~\cite{Horowitz:2011dz,Wang:2011rva,Siani:2011uj,Wang:2011ri,Wang:2012yj} that for the asymptotic AdS geometry the relation between the condensate within the barrier at zero current $\langle \mathcal{O} \rangle_{x=0}$ and the width of the junction $\ell$, as well as the relation between the maximal current (or critical current) $J_{max}$ and the width of the junction $\ell$ behave as
\be
\label{fitrelationo}\langle \mathcal{O}\rangle_{x=0}/T_c^{3/z}&\approx& A_1 e^{-\frac{\ell}{2 \xi}},\\
\label{fitrelationj}J_{max}/T_c^{(1+z)/z}&\approx& A_0 e^{-\frac{\ell}{\xi}}.
\ee
Those behaviour is in good agreement with condensed matter physics~\cite{tinkham}, according to which $\xi$ is identified as the normal metal coherence length.
\begin{figure}[h!]
\centering
   \includegraphics[scale=0.258]{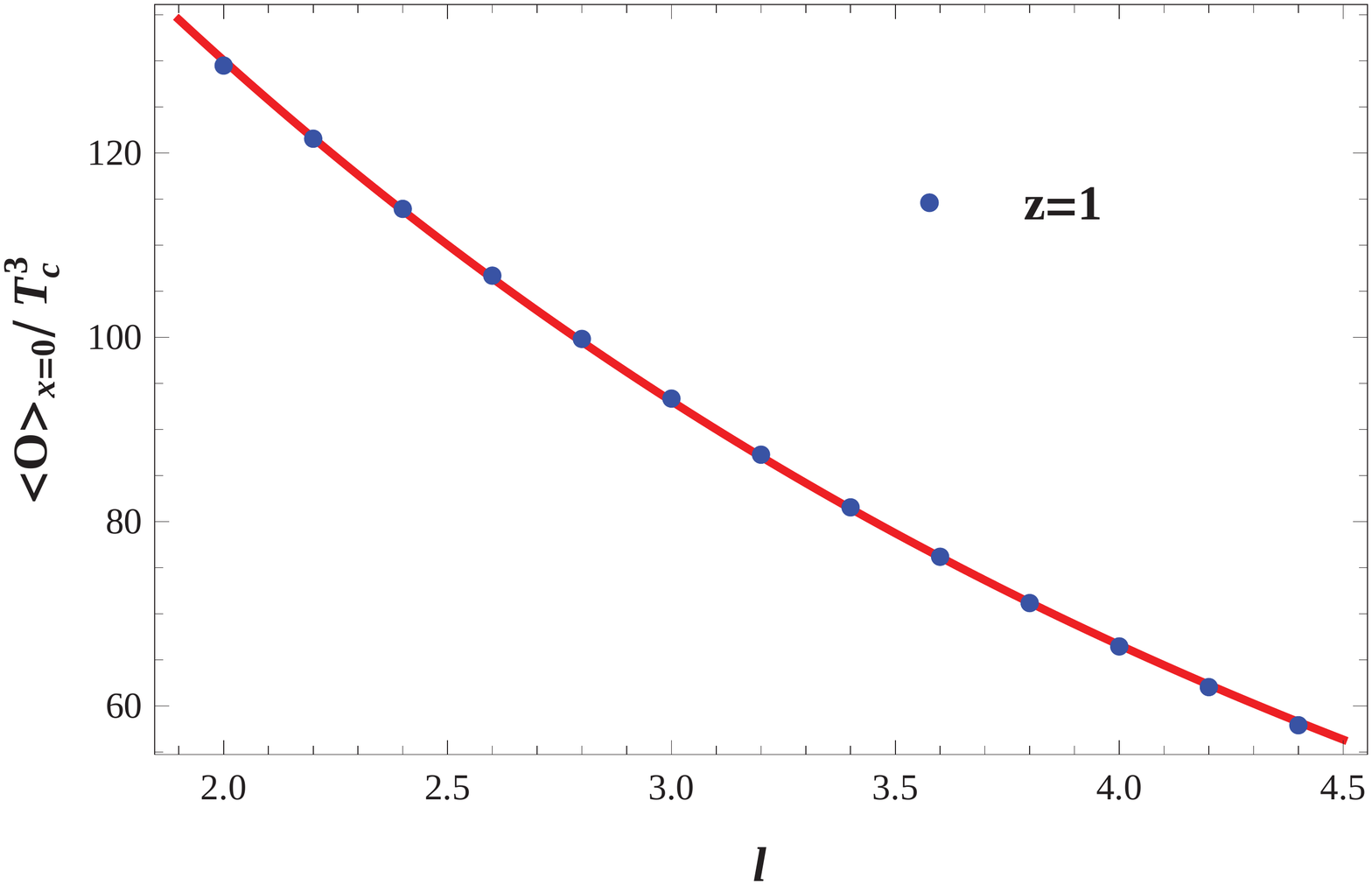}
    \includegraphics[scale=0.258]{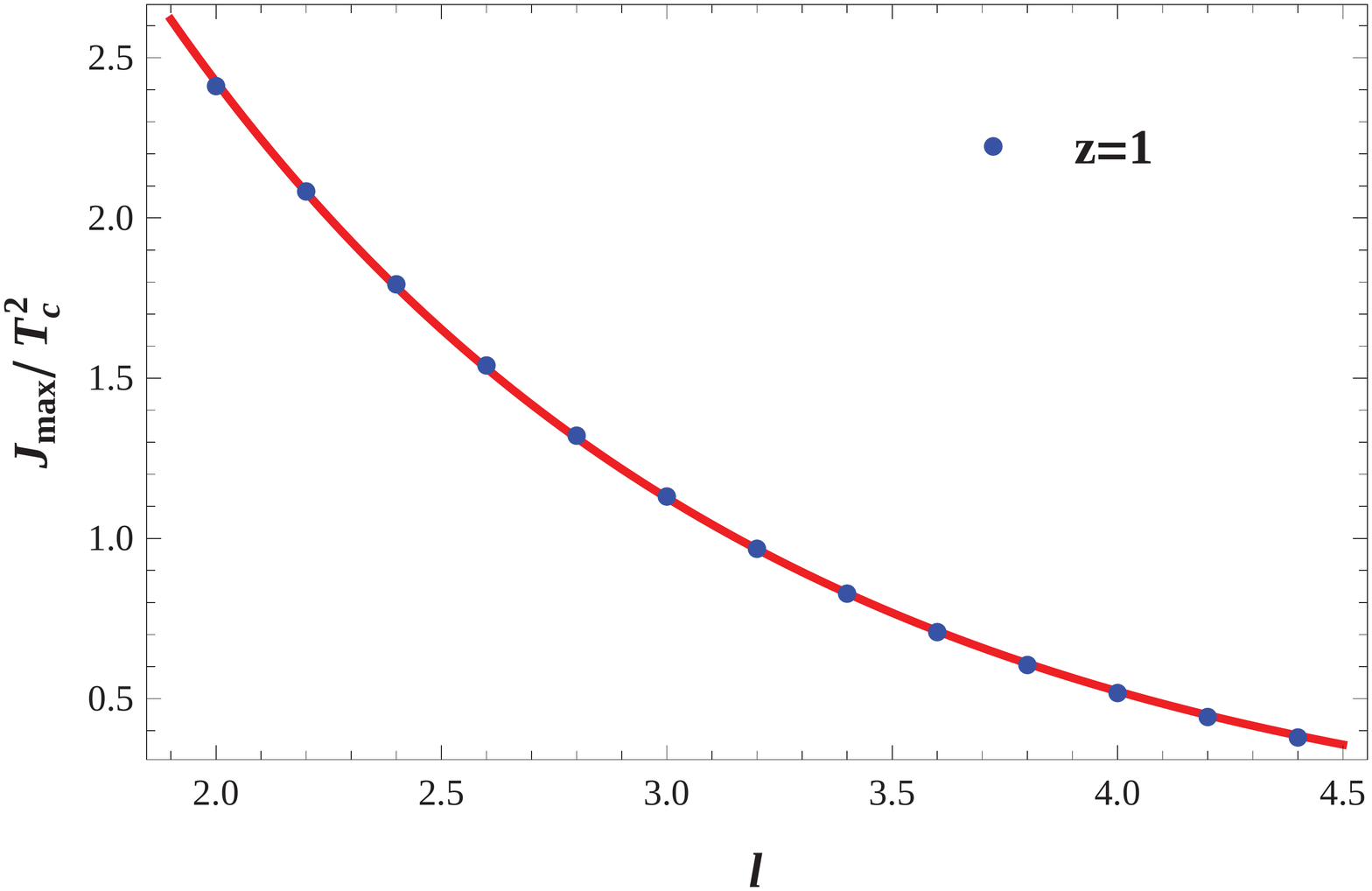}
          \caption{\label{OJL1}  $\langle \mathcal{O}\rangle_{x=0}/T_c^{3/z}$ (left) and $J_{max}/T_c^{(1+z)/z}$ (right) as functions of $\ell$ for $z=1$. The parameters are $\mu_\infty=10.5, \sigma=0.7, \epsilon=0.7$  and $2\leq\ell\leq 4.4$. The dots are from the numerics while the red lines are the fitted curves.  }
\end{figure}

We indeed reproduce similar results. On the left panel of figure.~\eqref{OJL1}, we plot the relation of $\langle O\rangle_{x=0}/T_c^{3/z}$ and $\ell$ for $z=1$ and find that they satisfy a decreasing exponential relation as
\begin{eqnarray}\label{fito1}
 \langle\mathcal{O}\rangle_{x=0}/T_c^{3} \approx 253.896 \times e^{\frac{-\ell}{2\times{ 1.49478}}}, \quad \text{for} \quad z=1.
\end{eqnarray}
The relation between $J_{max}/T_c^{(1+z)/z}$ and $\ell$ can be found on the right panel of figure.~\eqref{OJL1}. The fitting curve satisfy the following relation,
\begin{eqnarray}\label{fitj1}
J_{max}/T_c^{2}\approx 11.2449\times e^{\frac{-\ell}{1.30389}}, \quad \text{for} \quad z=1.
\end{eqnarray}
We can find that for $z=1$ the fitted value of $\xi$ from the two relations~\eqref{fito1} and~\eqref{fitj1} are very close to each other, with the error of $12.77\%$.~\footnote{\label{foot} Due to the limitation of numerics, one can not choose a too steep profile for $\mu(x)$. Thus the normal and superconducting phases in the junction are not cleanly separated. This is argued to justify the disagreement between the two estimates of $\xi$~\cite{Horowitz:2011dz}.}

\subsection{The case of $z=2$}
For $z=2$,  the  asymptotic expansion of $M_t$ near the boundary is $M_t\sim\rho(x)-\mu(x)\log(r)$. For convenience of the numerical calculation, we make a transformation
\be M_t\to \frac{\log(r)}{1-1/r}M_t,\ee
 The reason for dividing $(1-1/r)$ in the denominator is that at the horizon $r_0=1$, we need to impose the coefficients ${\log(r)}/({1-1/r})$ be non-vanishing, thus the new fields $M_t$ at horizon can have a specific vanishing boundary condition. This step of scaling $M_t$ is crucial for the numerics, and we find it is much feasible for the codes.
 \begin{figure}[h]
\centering
    \includegraphics[scale=0.35]{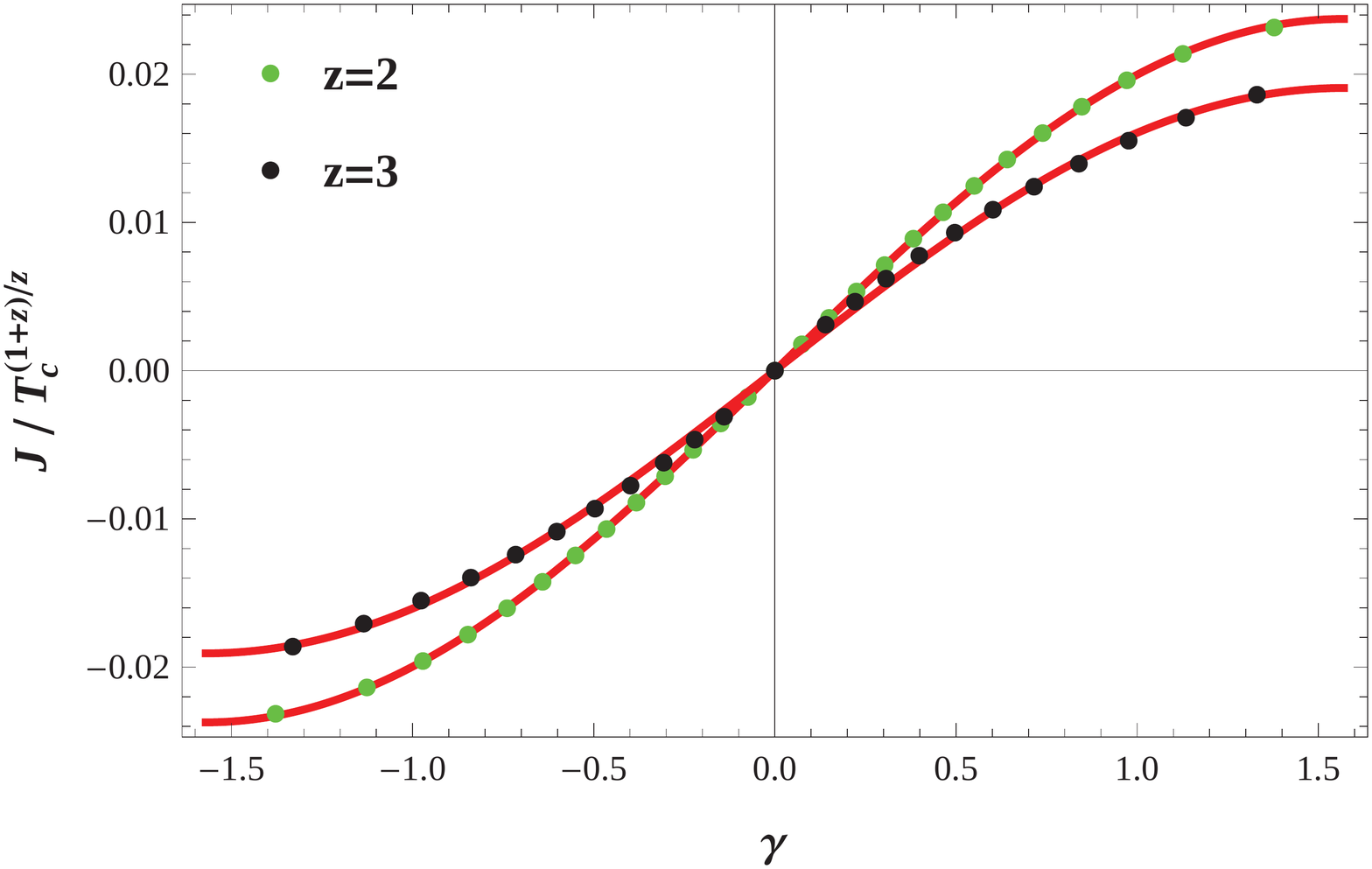}
      \caption{\label{Jgamma23} The behaviour of $J/T_c^{(1+z)/z}$ as a function of $\gamma$ for $z=2$ (green dots) and $z=3$ (black dots).  The parameters we use are $\mu_\infty=10.5, \ell=3, \sigma=0.7$ and $\epsilon=0.7$. The dots are from the numerics while the the red lines are the best fit sin curves of these dots. }
\end{figure}

The relation between the current and the phase difference can be found in figure.~\eqref{Jgamma23} in which the green dots are from the numerics while the red line is the best fitted curve. In this case, the asymptotic behaviour is much more different from the previous one. However, we find that the famous sinusoidal relation between current and phase difference is still satisfied very well. The numerical calculation shows that
 \be
 J/T_c^{3/2}\approx 0.02372\sin(\gamma), \quad \text{for}\quad z=2.
 \ee

\begin{figure}[h]
\centering
    \includegraphics[scale=0.35]{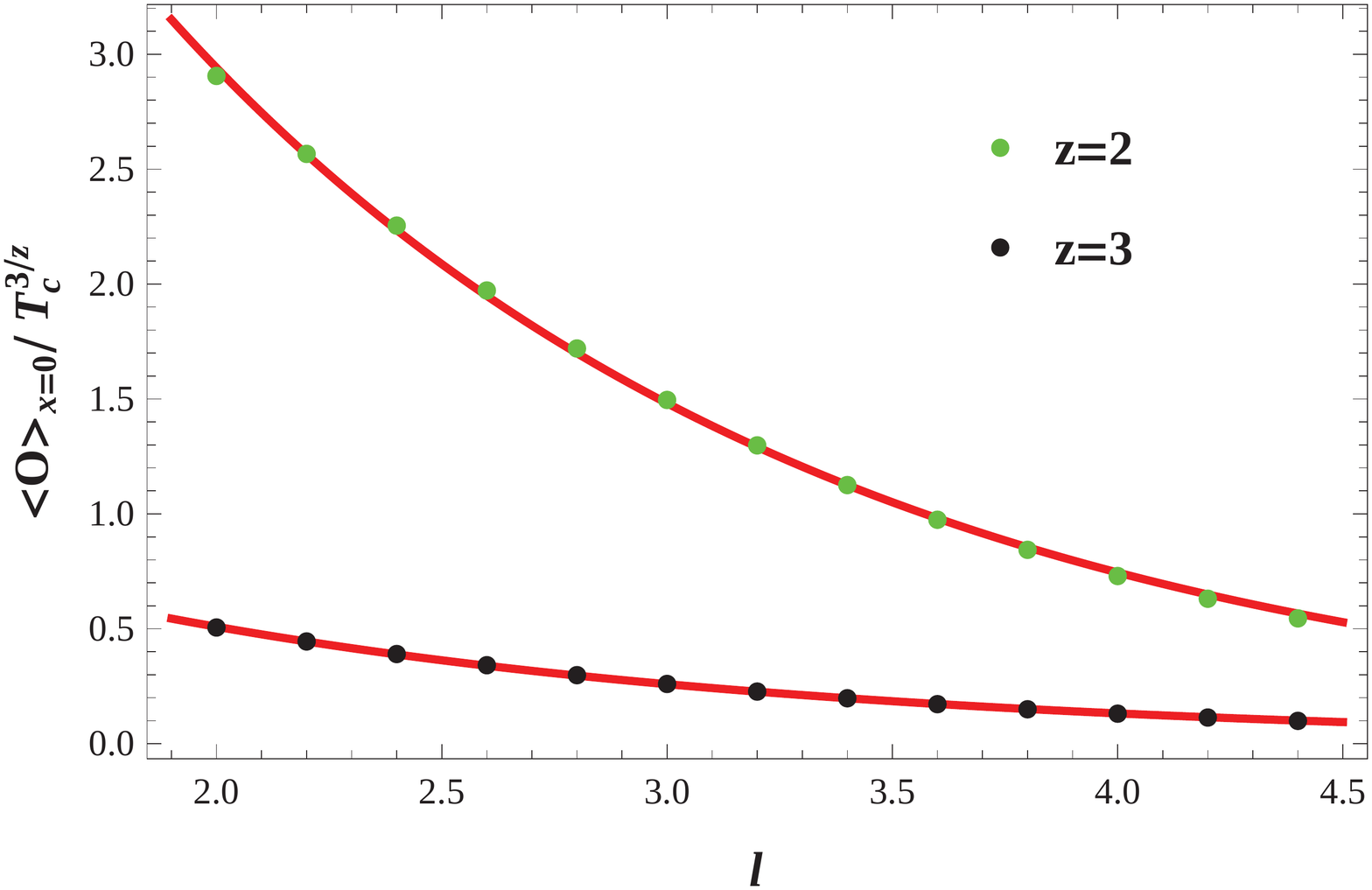}
          \caption{\label{OL23}  Relations between $\langle \mathcal{O}\rangle_{x=0}/T_c^{3/z}$ at zero current and $\ell$ for Lifshitz scaling. The upper curve and the lower curve correspond to $z=2$ and $z=3$, respectively. In both cases, the points are from numerics and red lines are fitted curves. We choose $\mu_\infty=10.5, \sigma=0.7$ and $\epsilon=0.7$.  }
\end{figure}

Let us consider the behaviour of the condensate at the centre of the contact at zero current. As one can see in figure~\eqref{OL23} that the condensate as a function of the length of the link can be fitted very well by the exponential decreasing function, which reads
\be
\label{olz2}\langle \mathcal{O}\rangle_{x=0}/T_c^{3/2}\approx 11.6122 \times e^{-\frac{\ell}{2\xi_o}},\   \ \xi_o\approx0.72835\quad \text{for} \quad z=2.
\ee
Compared to the relation~\eqref{fitrelationo}, maybe with a little abuse of terminology, this result encourages us to identify $\xi_o$ as the coherence length.~\footnote{A priori, Lifshitz case would be different form its AdS counterpart. To stress this issue, we use $\xi_o$ and $\xi_j$ to denote different coherence lengths extracted form the condensate and critical current, respectively. }

\begin{figure}[h]
\centering
   \includegraphics[scale=0.35]{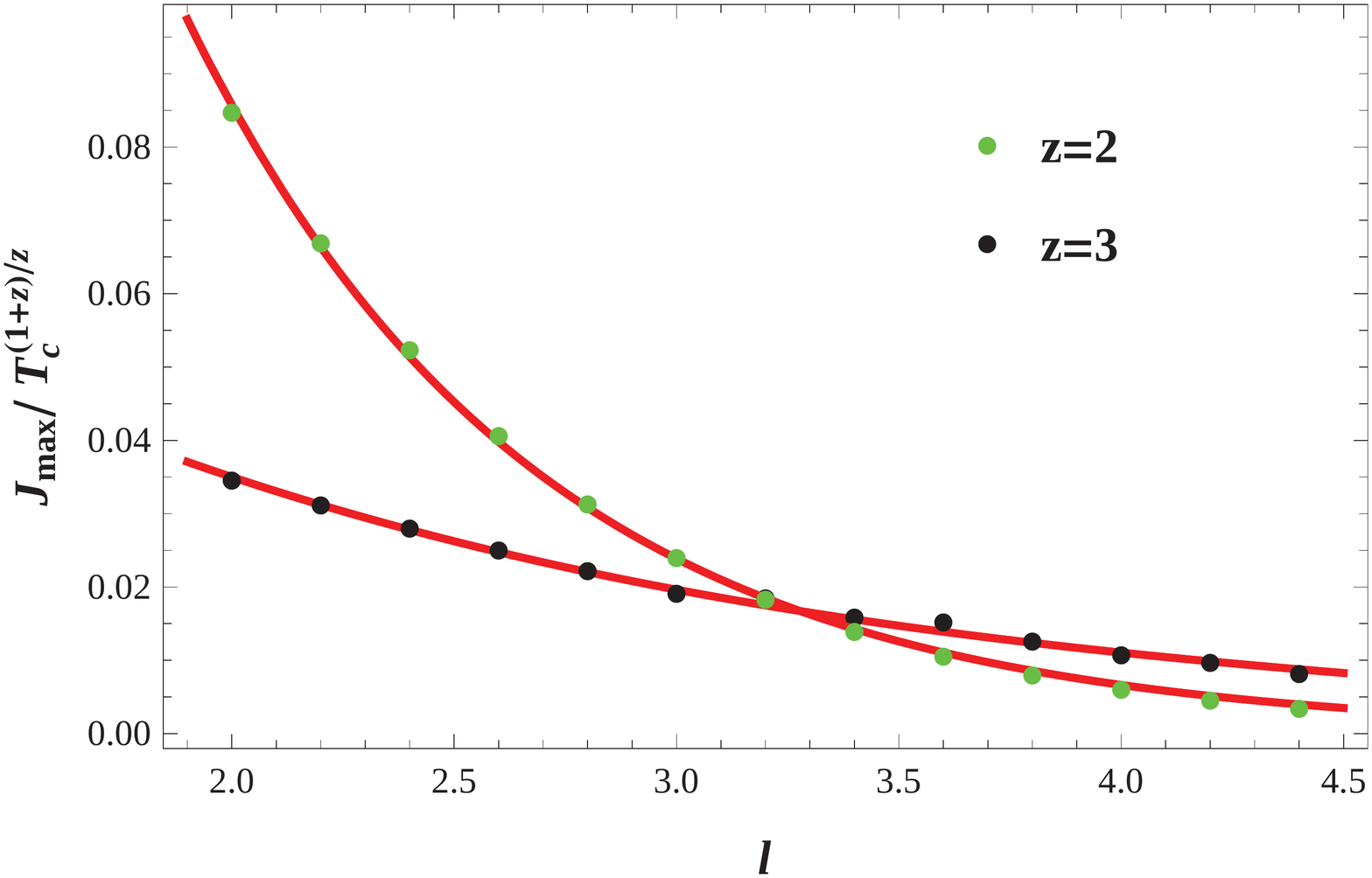}
      \caption{\label{JmaxL23}  Relations between $J_{max}/T_c^{(1+z)/z}$ and $\ell$ for Lifshitz scaling. The upper curve and the lower curve correspond to $z=2$ and $z=3$, respectively. In both cases, the points come from numerics and red lines are best fit curves. We use $\mu_\infty=10.5, \sigma=0.7$ and $\epsilon=0.7$ in the plot. }
\end{figure}

The dependence of $J_{max}$ on $\ell$ is shown in figure~\eqref{JmaxL23}. Once again, this cure can be fitted by an the exponential decreasing relation
\be
\label{jlz2}J_{max}/T_c^{3/2}\approx 1.10613\times e^{-\frac{\ell}{\xi_j}},\   \ \xi_j\approx 0.782017\quad \text{for} \quad z=2.
\ee
Comparing to~\eqref{fitrelationj}, one may also consider $\xi_j$ as the coherence length. We can see that the discrepancy between the value of coherence length $\xi$ obtained from~\eqref{olz2} and~\eqref{jlz2} is consistent with each other within the error $6.9\%$.

\subsection{The case of $z=3$}
For $z=3$,  the  asymptotic expansion of $M_t$ near the boundary is $M_t\sim\mu(x)-\rho(x)r$.
In this case, we introduce a transformation
\be M_t\to r M_t, \ee
in our numerical calculation.
Note that although now $\mu(x)$ is the subleading term in the expansion, we can still regard it as the chemical potential according to the explanation in ref.~\cite{Hartnoll:2009ns}.

The relation between the current and the phase difference is again shown in figure.~\eqref{Jgamma23}, where the black dots are from the numerics while the red line is the best fitted curve using sinusoidal function. It satisfies the relation as
\be
J/T_c^{4/3}\approx 0.01906\sin(\gamma),\quad \text{for} \quad z=3.
\ee
From above relation one can read off the maximum current $J_{max}\approx0.01906$ for the choosing parameters.

Meanwhile, the dependence of  $\langle \mathcal{O} \rangle_{x=0}$ at zero current on $\ell$ as well as $J_{max}$ on $\ell$ can be found in figure.~\eqref{OL23} and figure.~\eqref{JmaxL23}, respectively. Both can be fitted very well by exponential decreasing functions and the final results read
\be
\langle \mathcal{O}\rangle_{x=0}/T_c\approx 1.9632 \times e^{-\frac{\ell}{2 \xi_o}},\   \ \xi_{o}\approx0.740657\quad \text{for} \quad z=3,\\
J_{max}/T_c^{4/3}\approx 0.110907\times e^{-\frac{\ell}{\xi_j}},\   \ \xi_j\approx1.73365 \quad \text{for} \quad z=3.
\ee
Surprisingly, we see that the values of $\xi_o$ and $\xi_j$ exhibit enormous discrepancy. This large discrepancy can neither be explained by numerical error, nor by the disagreement corresponding to the profile of $\mu(x)$ as commented in footnote~\eqref{foot}.  We shall discuss this issue in the next section.

\section{Conclusion and Discussion}
\label{sect:conclusion}
In this work, we holographically studied the properties of SNS Josephson junction in non-relativistic case with Lifshitz scaling.
It can be carried out in terms of the Abelian-Higgs model~\cite{Hartnoll:2008vx} coupled with an asymptotic Lifshitz black brane solution in gravity side. Due to the presence of the {\it dynamical critical exponent} $z$, the asymptotic expansions of the fields behave distinctly from each other for different $z$. Therefore, it was expected that the properties of the Josephson junctions would depend on $z$ as well.

By virtue of the Chebyshev spectral methods, we could solve the coupled PDEs~\eqref{pdes} successfully. We found that the famous sinusoidal relations between the current $J$ and phase difference $\gamma$  across the weak link still exists for various $z$. Furthermore, our results showed that there is indeed an exponential decreasing relation between the condensate $\langle \mathcal{O}\rangle$ at the middle of the barrier with vanishing current and the width of the link $\ell$. Similar relation also holds  between the critical current $J_{max}$ and $\ell$.  As a consistent check for our numerics, let us consider the behaviour of the coefficient $A_1$ form the relation~\eqref{fitrelationo} with respect to $z$.  Note that we set the same profile of the chemical potential $\mu(x)$ for various $z$ and the critical chemical potentials $\mu_c$ for a homogeneous superconductor would increase with $z$ increased. Therefore, the chemical potential at $x=0$ in the link would be much smaller compared to $\mu_c$ if $z$ is much larger. Hence the condensate $\langle \mathcal{O}\rangle$ at $x=0$ would be much smaller as $z$ increases. This would in turn made $A_1$ smaller when $z$ becomes bigger. This is nothing but we found in our numerical calculation.

In order to compare with the relativistic case $z=1$, we also reproduced the holographic junction in AdS-Schwarzschild black brane. Similar to $z=1$ case, we found that for $z=2$ the coherence length $\xi$ obtained from the condensate within the link (see equation~\eqref{olz2}) and the one from the critical current  (see equation~\eqref{jlz2}) were close to each other within acceptable error. However, for $z=3$ the $\xi$'s got from the two relations were no longer consistent. Although we calculated the case of $z=3$ by using much higher precision, we still could not render the two $\xi$'s consistent.  One should keep in mind that the relations~\eqref{olz2} and~\eqref{jlz2} are deduced form conventional superconductivity under some additional approximations.
In contrast, our holographic construction is, in principle, only applicable for the superconductors at strong coupling, thus, a priori, would far deviate form the conventional one which is weakly interacted.  An instructive example is to consider the well-known Abelian-Higgs model in AdS case. As the temperature decreases, there exists a gap frequency $\omega_g$ from the optical conductivity in the superconducting phase, and one can also read off the energy gap $\Delta_g$ from the low temperature behaviour of normal contribution to the DC conductivity. In a standard weak coupling picture of superconductivity, the gap $\omega_g$ is understood as the energy required to break a Cooper pair into its constitutive electrons and the energy of
the constituent quasiparticles is given by $\Delta_g$. In BCS theory $\omega_g=2\Delta_g$, while it does not hold in holographic setup~\cite{Horowitz:2008bn, Hartnoll:2008kx}, indicating that we are clearly not in a weak coupling regime and that such a quasiparticle picture is not applicable. Therefore, we expect that our results for Josephson junctions with Lifshitz scaling may suggest new mechanism compared to AdS case. It is desirable to understand our results form condensed matter theory point view. It will be interesting to see whether one can construct, for example, a Landau-Ginzberg like theory with Lifshitz scaling that exhibits similar deviation in this paper. 

Similar discussion can be straightforwardly generalized to include hyperscaling violation characterised by $\theta$, another important exponent in low energy physics of condensed matter system. In cases with general $\theta$ and $z$, the dual non-relativistic theory is not even scale invariant, qualitatively different from its Lifshitz counterpart. Nevertheless, we expect that the main results would be similar to the Lifshitz geometry. There are various kinds of junctions, and the properties of these junctions can be considerably different. It is, however, known that a sinusoidal current-phase relation is only a special case in Josephson tunneling, which is attainable only for  such as temperatures sufficiently close to critical temperature or sufficiently high and wide potential barriers between two superconductors~\cite{ Likharev,Arnold}. So far, similar studies initiated from ref.~\cite{Horowitz:2011dz} all produced the sinusoidal relation between current and phase difference, including the case with Lifshitz scaling in current paper. To obtain the non-sinusoidal relation, one is suggested to consider the case with much lower temperatures. Therefore, it is natural to include the back reaction of matter fields to the geometry~\cite{wang2014}. It will be also interesting to extend similar study to other types of junctions and to cases with competing orders. We hope to report related issues in the future.

\section*{Acknowledgements}

 The authors are grateful to Rong-Gen Cai for his valuable discussions, and the hospitality and partial support from the Kavli Institute for Theoretical Physics China (KITPC) during the completion of this work. HFL was supported  in part by the Young Scientists Fund of the National Natural Science Foundation of China (No.11205097), the Program for the Innovative Talents of Higher Learning Institutions of Shanxi, the Natural Science Foundation for Young Scientists of Shanxi Province, China (No.2012021003-4) and the Shanxi Datong University doctoral Sustentation Fund(No.2011-B-04), China;  LL was supported in part by European Union's Seventh Framework Programme under grant agreements (FP7-REGPOT-2012-2013-1) no 316165, the EU-Greece program ``Thales" MIS 375734 and was also co-financed by the European Union (European Social Fund, ESF) and Greek national funds through the Operational Program ``Education and Lifelong Learning" of the National Strategic Reference Framework (NSRF) under ``Funding of proposals that have received a positive evaluation in the 3rd and 4th Call of ERC Grant Schemes";  HQZ was partially supported by a Marie Curie International Reintegration Grant PIRG07-GA-2010-268172 and the Young Scientists Fund of the National Natural Science Foundation of China (No.11205097).
\appendix

\end{CJK*}

\begin{thebibliography}{99}



\baselineskip 12pt

\bibitem{Maldacena:1997re}
  J.~M.~Maldacena,
  ``The large N limit of superconformal field theories and supergravity,''
  Adv.\ Theor.\ Math.\ Phys.\  {\bf 2}, 231 (1998)
  [Int.\ J.\ Theor.\ Phys.\  {\bf 38}, 1113 (1999)]
  [arXiv:hep-th/9711200].
\bibitem{Gubser:1998bc}
  S.~S.~Gubser, I.~R.~Klebanov and A.~M.~Polyakov,
  ``Gauge theory correlators from non-critical string theory,''
  Phys.\ Lett.\  B {\bf 428}, 105 (1998)
  [arXiv:hep-th/9802109].
\bibitem{Witten:1998qj}
  E.~Witten,
  ``Anti-de Sitter space and holography,''
  Adv.\ Theor.\ Math.\ Phys.\  {\bf 2}, 253 (1998)
  [arXiv:hep-th/9802150].

\bibitem{Hartnoll:2008vx}
  S.~A.~Hartnoll, C.~P.~Herzog and G.~T.~Horowitz,
  ``Building a Holographic Superconductor,''
  Phys.\ Rev.\ Lett.\  {\bf 101}, 031601 (2008)
  [arXiv:0803.3295 [hep-th]].

  \bibitem{Gubser:2008wv}
  S.~S.~Gubser and S.~S.~Pufu,
  ``The Gravity dual of a p-wave superconductor,''
  JHEP {\bf 0811}, 033 (2008)
  [arXiv:0805.2960 [hep-th]].

\bibitem{Aprile:2010ge}
  F.~Aprile, D.~Rodriguez-Gomez and J.~G.~Russo,
  ``p-wave Holographic Superconductors and five-dimensional gauged Supergravity,''
  JHEP {\bf 1101}, 056 (2011)
  [arXiv:1011.2172 [hep-th]].

\bibitem{Cai:2013aca}
  R.~G.~Cai, L.~Li and L.~F.~Li,
  ``A Holographic P-wave Superconductor Model,''
  JHEP {\bf 1401}, 032 (2014)
  [arXiv:1309.4877 [hep-th]].

\bibitem{Chen:2010mk}
  J.~W.~Chen, Y.~J.~Kao, D.~Maity, W.~Y.~Wen and C.~P.~Yeh,
  ``Towards A Holographic Model of D-Wave Superconductors,''
  Phys.\ Rev.\ D {\bf 81}, 106008 (2010)
  [arXiv:1003.2991 [hep-th]].

\bibitem{Benini:2010pr}
  F.~Benini, C.~P.~Herzog, R.~Rahman and A.~Yarom,
  ``Gauge gravity duality for d-wave superconductors: prospects and challenges,''
  JHEP {\bf 1011}, 137 (2010)
  [arXiv:1007.1981 [hep-th]].

\bibitem{Herzog:2009xv}
  C.~P.~Herzog,
  ``Lectures on Holographic Superfluidity and Superconductivity,''
  J.\ Phys.\ A {\bf 42}, 343001 (2009)
  [arXiv:0904.1975 [hep-th]].

\bibitem{Horowitz:2010gk}
  G.~T.~Horowitz,
  ``Introduction to Holographic Superconductors,''
  Lect.\ Notes Phys.\  {\bf 828}, 313 (2011)
  [arXiv:1002.1722 [hep-th]].

\bibitem{Musso:2014efa}
  D.~Musso,
  ``Introductory notes on holographic superconductors,''
  arXiv:1401.1504 [hep-th].
  
\bibitem{Domenech:2010nf} 
  O.~Domenech, M.~Montull, A.~Pomarol, A.~Salvio and P.~J.~Silva,
  ``Emergent Gauge Fields in Holographic Superconductors,''
  JHEP {\bf 1008}, 033 (2010)
  [arXiv:1005.1776 [hep-th]].

\bibitem{Horowitz:2011dz}
G.~T.~Horowitz, J.~E.~Santos and B.~Way,
``A Holographic Josephson Junction,''  Phys.\ Rev.\ Lett.\  {\bf 106}, 221601 (2011)  [arXiv:1101.3326 [hep-th]].

\bibitem{Wang:2011rva}
  Y.~Q.~Wang, Y.~X.~Liu and Z.~H.~Zhao,
  ``Holographic Josephson Junction in 3+1 dimensions,''
  arXiv:1104.4303 [hep-th].

\bibitem{Siani:2011uj}
  M.~Siani,
  ``On inhomogeneous holographic superconductors,''
  arXiv:1104.4463 [hep-th].

\bibitem{Wang:2011ri}
  Y.~Q.~Wang, Y.~X.~Liu and Z.~H.~Zhao,
  ``Holographic p-wave Josephson junction,''
  arXiv:1109.4426 [hep-th].

\bibitem{Wang:2012yj}
  Y.~Q.~Wang, Y.~X.~Liu, R.~G.~Cai, S.~Takeuchi and H.~Q.~Zhang,
  ``Holographic SIS Josephson Junction,''
  JHEP {\bf 1209}, 058 (2012)
  [arXiv:1205.4406 [hep-th]].


\bibitem{Rozali:2013pla}
  M.~Rozali and A.~Vincart-Emard,
  ``Chiral Edge Currents in a Holographic Josephson Junction,''
  JHEP {\bf 1401}, 003 (2014)
  [arXiv:1310.4510 [hep-th]].
 

\bibitem{Cai:2013sua}
  R.~G.~Cai, Y.~Q.~Wang and H.~Q.~Zhang,
  ``A holographic model of SQUID,''
  JHEP {\bf 1401}, 039 (2014)
  [arXiv:1308.5088 [hep-th]].
  
\bibitem{Takeuchi:2013kra} 
  S.~Takeuchi,
  ``Holographic Superconducting Quantum Interference Device,''
  arXiv:1309.5641 [hep-th].
  
  
\bibitem{Kiritsis:2011zq}
  E.~Kiritsis and V.~Niarchos,
  ``Josephson Junctions and AdS/CFT Networks,''
  JHEP {\bf 1107}, 112 (2011)
  [Erratum-ibid.\  {\bf 1110}, 095 (2011)]
  [arXiv:1105.6100 [hep-th]].

\bibitem{Domokos:2012rj}
  S.~K.~Domokos, C.~Hoyos and J.~Sonnenschein,
  ``Holographic Josephson Junctions and Berry holonomy from D-branes,''
  JHEP {\bf 1210}, 073 (2012)
  [arXiv:1207.2182 [hep-th]].


\bibitem{Hartnoll:2009sz}
  S.~A.~Hartnoll,
  ``Lectures on holographic methods for condensed matter physics,''
  Class.\ Quant.\ Grav.\  {\bf 26}, 224002 (2009)
  [arXiv:0903.3246 [hep-th]].

\bibitem{Sachdev:2010ch}
  S.~Sachdev,
  ``Condensed Matter and AdS/CFT,''
  Lect.\ Notes Phys.\  {\bf 828}, 273 (2011)
  [arXiv:1002.2947 [hep-th]].

\bibitem{Hartnoll:2009ns}
S.~A.~Hartnoll, J.~Polchinski, E.~Silverstein and D.~Tong,
``Towards strange metallic holography,''  JHEP {\bf 1004}, 120 (2010)  [arXiv:0912.1061 [hep-th]].

\bibitem{Lee:2010uy}
  B.~H.~Lee and D.~W.~Pang,
  ``Notes on Properties of Holographic Strange Metals,''
  Phys.\ Rev.\ D {\bf 82}, 104011 (2010)
  [arXiv:1006.4915 [hep-th]].

\bibitem{Kim:2010zq}
  B.~S.~Kim, E.~Kiritsis and C.~Panagopoulos,
  ``Holographic quantum criticality and strange metal transport,''
  New J.\ Phys.\  {\bf 14}, 043045 (2012)
  [arXiv:1012.3464 [cond-mat.str-el]].

\bibitem{Keranen:2011xs}
  V.~Keranen, E.~Keski-Vakkuri and L.~Thorlacius,
  ``Thermalization and entanglement following a non-relativistic holographic quench,''
  Phys.\ Rev.\ D {\bf 85}, 026005 (2012)
  [arXiv:1110.5035 [hep-th]].

\bibitem{Gursoy:2011gz}
  U.~Gursoy, E.~Plauschinn, H.~Stoof and S.~Vandoren,
  ``Holography and ARPES Sum-Rules,''
  JHEP {\bf 1205}, 018 (2012)
  [arXiv:1112.5074 [hep-th]].

\bibitem{Fang:2012pw}
  L.~Q.~Fang, X.~H.~Ge and X.~M.~Kuang,
  ``Holographic fermions in charged Lifshitz theory,''
  Phys.\ Rev.\ D {\bf 86}, 105037 (2012)
  [arXiv:1201.3832 [hep-th]].

\bibitem{Wu:2013xta}
  J.~P.~Wu,
  ``Holographic fermions on a charged Lifshitz background from Einstein-Dilaton-Maxwell model,''
  JHEP {\bf 1303}, 083 (2013).

\bibitem{Taylor:2008tg}
  M.~Taylor,
  ``Non-relativistic holography,''
  arXiv:0812.0530 [hep-th].


\bibitem{Brynjolfsson:2009ct}
  E.~J.~Brynjolfsson, U.~H.~Danielsson, L.~Thorlacius and T.~Zingg,
  ``Holographic Superconductors with Lifshitz Scaling,''
  J.\ Phys.\ A {\bf 43}, 065401 (2010)
  [arXiv:0908.2611 [hep-th]].

\bibitem{Sin:2009wi}
  S.~J.~Sin, S.~S.~Xu and Y.~Zhou,
  ``Holographic Superconductor for a Lifshitz fixed point,''
  Int.\ J.\ Mod.\ Phys.\ A {\bf 26}, 4617 (2011)
  [arXiv:0909.4857 [hep-th]].

\bibitem{Bu:2012zzb}
  Y.~Bu,
  ``Holographic superconductors with $z=2$ Lifshitz scaling,''
  Phys.\ Rev.\ D {\bf 86}, 046007 (2012)
  [arXiv:1211.0037 [hep-th]].

\bibitem{Lu:2013tza}
  J.~W.~Lu, Y.~B.~Wu, P.~Qian, Y.~Y.~Zhao and X.~Zhang,
  ``Lifshitz Scaling Effects on Holographic Superconductors,''
  Nucl.\ Phys.\ B {\bf 887}, 112 (2014)
  [arXiv:1311.2699 [hep-th]].

\bibitem{trefensen} L.~N.~Trefethen, Spectral methods in MATLAB, Siam, Philadelphia, (2000).

\bibitem{tinkham} M.~Tinkham, Introduction to Superconductivity, McGraw Hill, (1975).

\bibitem{Horowitz:2008bn}
  G.~T.~Horowitz and M.~M.~Roberts,
  ``Holographic Superconductors with Various Condensates,''
  Phys.\ Rev.\ D {\bf 78}, 126008 (2008)
  [arXiv:0810.1077 [hep-th]].

\bibitem{Hartnoll:2008kx}
  S.~A.~Hartnoll, C.~P.~Herzog and G.~T.~Horowitz,
  ``Holographic Superconductors,''
  JHEP {\bf 0812}, 015 (2008)
  [arXiv:0810.1563 [hep-th]].

\bibitem{Likharev}
 K.~K.~Likharev,
  ``Superconducting weak link,''
  Rev.\ Mod.\ Phys {\bf 51}, 101 (1979).

\bibitem{Arnold}
 G.~B.~Arnold,
  ``Superconducting tunneling without the tunneling Hamiltonian,''
 J.\ Low-Temp.\ Phys. {\bf 59}, 143 (1985).


\bibitem{wang2014}
R.~G.~Cai, L.~Li, Y.~X.~Liu, Y.~Q.~Wang and H.~Q.~Zhang,
  ``Holographic S-wave Josephson Junction with Back-Reaction.''
  (in progress)



\end{thebibliography}
\end{document}